\documentclass[pra,twocolumn,aps,showpacs,floatfix, nofootinbib]{revtex4-1}
\usepackage{amsmath}
\usepackage{amssymb,ams fonts,times,natbib}
\usepackage{graphicx}
\usepackage{epstopdf}

\begin{document}

\title{Manipulation of two-photon fluorescence spectra of chromophore aggregates with entangled photons; a simulation study} 
\author{Frank Schlawin}
\author{Konstantin E. Dorfman}
\author{Benjamin P. Fingerhut}
\author{Shaul Mukamel}
\email[]{smukamel@uci.edu}
\affiliation{Department of Chemistry, University of California, Irvine, California 92697-2025, USA}
\date{\today}
\pacs{123}

\begin{abstract}
The non-classical spectral and temporal features of entangled photons offer new possibilities to investigate the interactions of excitons in photosynthetic complexes, and to target the excitation of specific states. Simulations of fluorescence in the bacterial reaction center induced by entangled light demonstrate a degree of selectivity of double-exciton states which is not possible using classical stochastic light with the same power spectrum.
\end{abstract}

\maketitle

\section{Introduction}

Apart from their evident importance in experimental tests of the foundations of quantum mechanics, entangled photons promise many applications to quantum information processing \cite{Horodecki, Raymer2}, secure quantum communication \cite{Zeilinger, Braunstein, Braunstein2}, lithography \cite{lithography1, Sean, lithography2, Konstantin} or metrology \cite{metrology1, metrology2, Maccone}. In addition, their non-classical frequency and time correlations could also open up novel spectroscopic applications \cite{Richter1, Oleksiy2}. Entanglement-induced two-photon transparency \cite{Teich1} and the linear scaling of two-photon induced fluorescence with the pump intensity \cite{Dayan} constitute two basic non-classical effects observed with entangled photon pairs. This scaling makes it possible to carry out nonlinear optical measurements with much lower light intensity compared to classical light, as we will discuss in section \ref{lightsources}. More generally, entangled photons offer new control parameters for nonlinear spectroscopy \cite{Richter2}, and can be used to distinguish quantum pathways of matter \cite{Oleksiy1}. It remains however an open question, to what extent quantum entanglement is essential for these effects and whether some can be reproduced, for instance, by shaped or stochastic classical pulses \cite{Resch, Sergienko}. Recently, it has been argued that entanglement can induce collective resonances between non-interacting two-level atoms \cite{Scully1}. It was later found that these resonances may not be observed in fluorescence due to destructive interference of pathways \cite{Richter1}.\\
\begin{figure}[h!!]
 \centering
 \includegraphics[width=0.45\textwidth]{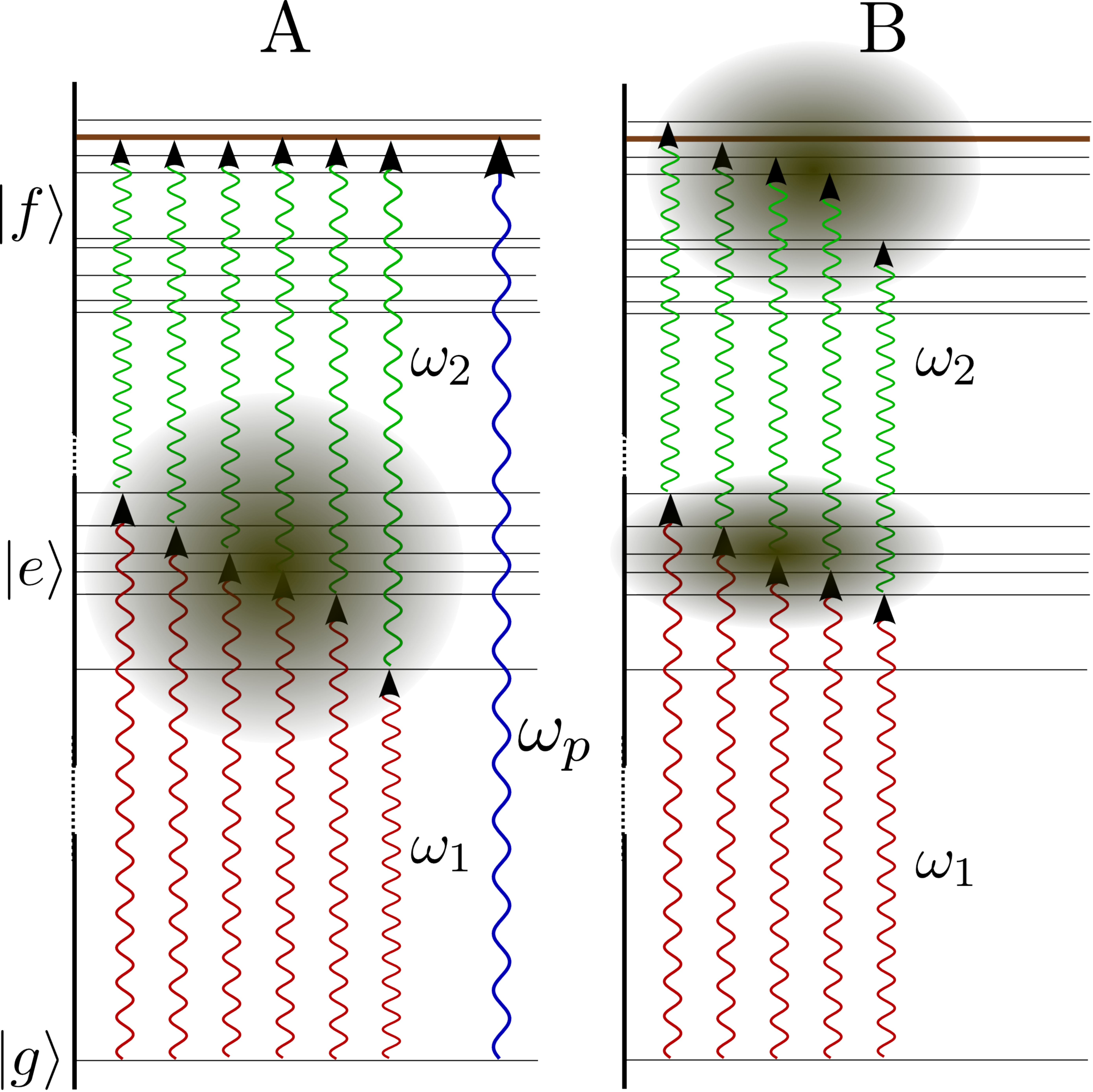}
\caption{(Color online) A: Two-photon transitions with entangled photons. The broad bandwidth allows us to access the entire manifold $\vert e \rangle$, at the same time the manifold $\vert f \rangle$ is well-resolved due to the narrow distribution of $\omega_1 + \omega_2$. B: Transitions with classical light. Bandwidths of $\omega_1$ and $\omega_2$ now add up, and the f-manifold cannot be resolved.}
 \label{transitions}
\end{figure}
In this Paper we present a theoretical framework for describing excitations induced by entangled photons in chromophore aggregates. These possess manifolds of single- and double-exciton states \cite{Kauffmann}, and the excitation of each double-exciton state can take place via many interfering pathways (see figure \ref{transitions}). We show that the non-classical spectral and temporal properties of twin photons can be used to excite state distributions that may not be obtained by classical light. Due to energy conservation in their generation, the sum of the energies of twin photons $\omega_p = \omega_1 + \omega_2$ has a sharp distribution ($i.e.$ $\Delta \omega_p$ is very small), even though each single photon may be broadband ($\Delta \omega_{1}, \Delta \omega_2$ very large). The high frequency resolution of $\omega_p$ allows to target certain double-exciton states, while the broad bandwidth of the individual beams enables the simultaneous access of all possible pathways that reach the double-exciton state (see figure \ref{transitions}). In contrast, with classical fields the bandwidths of the two beams add up, and create a large uncertainty in the sum. Our simulations show that the non-classical distributions could be detected by the fluorescence signal. The control parameters of the entangled light can be used to stretch the signal in two-dimensional plots, that could reveal additional information about the matter.

\section{The model}
\label{formalism}
We consider a chromophore aggregate interacting with the electromagnetic field and described by the Hamiltonian
\begin{equation}
H = H_0 + H_{F} + H_{\text{int}},
\end{equation}
$H_0$, $H_F$ and $H_{\text{int}}$ represent the aggregate, the field and their coupling, respectively. The electronic states group into well-separated manifolds, which are denoted as e- (single-exciton), f- (double-exciton) manifold and so on. The field couples to the system via the dipole operator, which induces transitions between these manifolds. In the rotating wave approximation, the interaction Hamiltonian reads
\begin{equation}
H_{\text{int}} (t) = V (t) E^{\dagger} (t) + V^{\dagger} (t) E (t) ,
\end{equation}
where we have introduced the positive-frequency component of the dipole operator 
\begin{equation}
V (t) = \sum_e \big( \mu_{eg} \vert g \rangle \langle e \vert e^{-i \omega_{eg} t} + \sum_f \mu_{fe} \vert e \rangle \langle f \vert e^{- i \omega_{fe} t}  \big),
\end{equation}
in which $\omega_{i j } = (E_i - E_j) / \hbar$ are matter transition frequencies, and $\mu_{ij}$ the dipole moments. The corresponding negative-frequency part of the electromagnetic field operator is given by
\begin{equation}
E^{\dagger} (t) = \sum_s \left( \frac{2 \pi \omega_s}{\Omega} \right)^{1 / 2} a^{\dagger}_s e^{i \omega_s t},
\end{equation}
where $\Omega$ denotes the quantization volume, $a^{\dagger}_s$ is the creation operator for mode s, and s runs over the relevant modes. 
Assuming perfect phase matching, we can neglect the spatial dependencies of the electromagnetic field. This is justified for large samples. 
Coupling with a phonon bath can induce many interesting relaxation effects. These are not included in the present study.
\begin{figure*}[ht!!]
 \centering
 \includegraphics[width=0.8\textwidth]{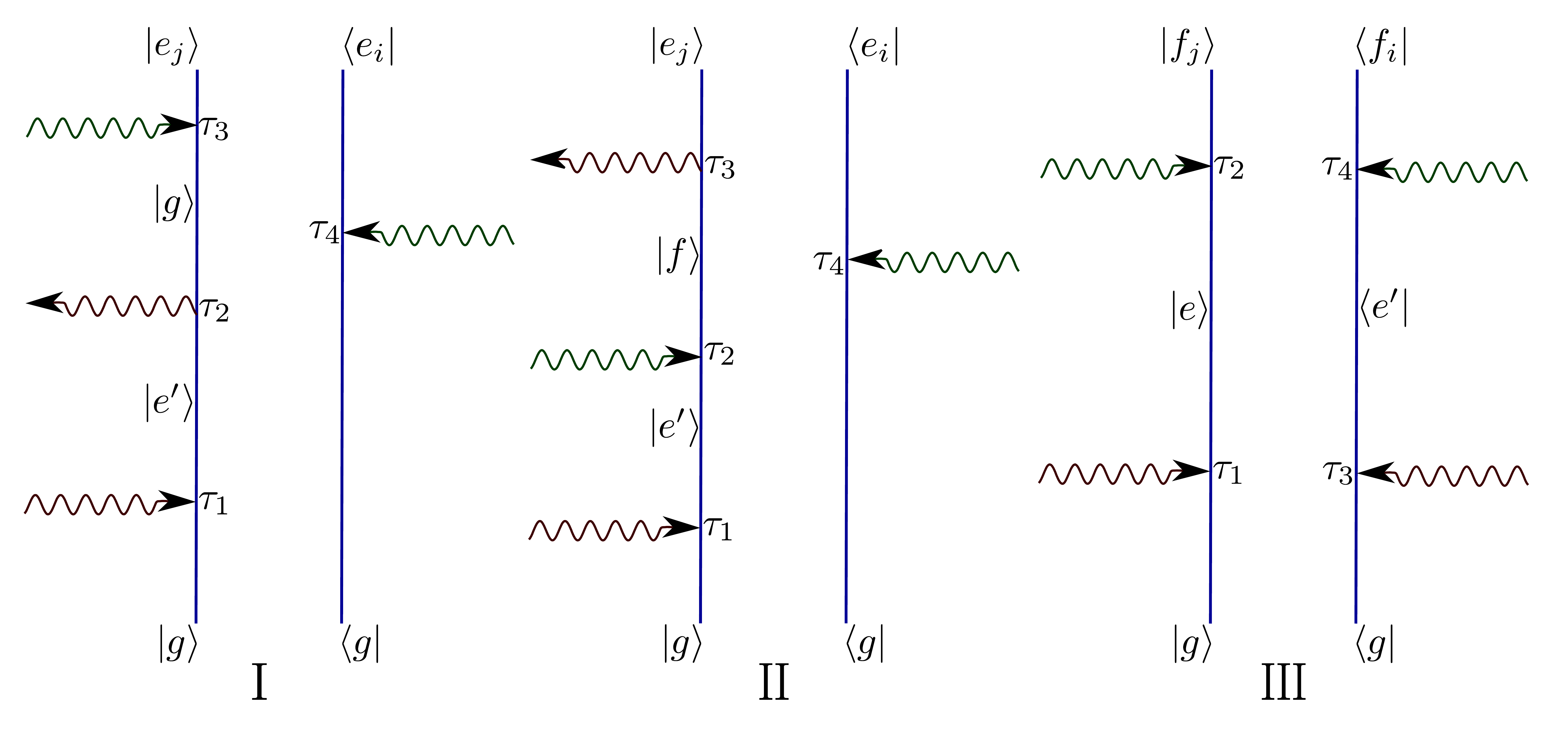}
 \caption{(Color online) Diagrams I and II: the single-exciton density matrix. Diagram III: the two-exciton density matrix.}
 \label{single2}
\end{figure*}
We first derive general expressions for density matrix elements in the single- and the double-exciton manifold in terms of convolutions of matter- and field-correlation functions. These results are valid for light with arbitrary statistical properties, and can describe two-photon induced fluorescence and related measurements in a consistent manner. 
The matrix element $(e_i, e_j)$ of the single-exciton part of the density matrix is given by
\begin{equation}
\varrho_{e_i e_j} (t) = \text{tr} [ \vert e_j (t) \rangle \langle e_i (t) \vert \varrho (t) ], \label{pe(t)}
\end{equation}
where $\varrho (t)$ denotes the density matrix of the entire matter-field system. In a superoperator Liouville space representation, the formal solution of the Heisenberg equation is given by the Dyson series \cite{Book}
\begin{equation}
\varrho (t) = \mathcal{T} \exp \left( - \frac{i}{\hbar} \int^{t} d\tau H_{\text{int}, -} (\tau) \right) \varrho (- \infty),
\end{equation}
where $\mathcal{T}$ is the time-ordering operator, and the superoperator $H_{\text{int}, -}$ is defined through $A_{-} \varrho = A \varrho - \varrho A$ in the interaction picture with respect to $H_0 + H_F$. For weak fields we can expand the exponential, and the leading (second-) order contribution to eq. (\ref{pe(t)}) yields
\begin{align}
 \varrho'_{e_i e_j} (t; \Gamma) &= - \left( - \frac{i}{\hbar} \right)^2 \int^t_{-\infty}\!\!\!\!\!\! d\tau_1 \int^{t}_{-\infty} \!\!\!\!\!\! d\tau_2 \notag \\
&\times \big\langle V (\tau_2) A_{ij} (t) V^{\dagger} (\tau_1) \rangle \langle E^{\dagger} (\tau_2) E (\tau_1) \big\rangle, \label{S2a} 
\end{align}
where $A_{ij} (t) \equiv  \vert e_j (t) \rangle \langle e_i (t) \vert$, and $\Gamma$ denotes the set of control parameters that govern the excitation, $e.g.$ frequencies, pulse envelopes etc. The square brackets $\langle \dots \rangle$ denote the quantum mechanical expectation value with respect to the initial state $\varrho (-\infty)$,
\begin{align}
\varrho (-\infty) &= \vert g \rangle \langle g \vert \otimes \varrho_{\text{field}},
\end{align}
$i.e.$ the aggregate is in the ground state $\vert g \rangle$, and the electromagnetic field is prepared in a state $\varrho_{\text{field}}$. \\
The fourth-order processes that end in the $\vert e \rangle$ manifold can be represented by the diagrams I and II shown in figure \ref{single2} (for diagram rules, see \cite{Rahav}). These can be divided into two classes: Raman type that only involve single-exciton resonances (diagram I), and two-photon absorption type (diagram II) that involve the double-exciton manifold. We have to take two additional diagrams into account, in which three interactions occur on the right-hand side. Apart from the projector $\vert e_j (t) \rangle \langle e_i (t) \vert$, these are simply the complex conjugates of diagrams I and II in figure \ref{single2}. We can thus write them as 
\begin{widetext}
\begin{align}
\varrho_{e_i e_j, \text{I}} (t; \Gamma) &= - \left( - \frac{i}{\hbar} \right)^4 \int^t_{-\infty}\!\!\!\!\!\! d\tau_4  \int^t_{-\infty}\!\!\!\!\!\! d\tau_3 \int^{\tau_3}_{-\infty}\!\!\!\! d\tau_2 \int^{\tau_2}_{-\infty}\!\!\!\!\!\! d\tau_1 \notag \\
&\times \bigg( \big\langle V (\tau_4) A_{ij} (t) V^{\dagger} (\tau_3) V (\tau_2) V^{\dagger} (\tau_1) \big\rangle \big\langle E^{\dagger} (\tau_4) E (\tau_3) E^{\dagger} (\tau_2) E (\tau_1) \big\rangle\notag \\
&+ \big\langle V (\tau_1) V^{\dagger} (\tau_2) V (\tau_3) A_{ij} (t) V^{\dagger} (\tau_4) \big\rangle \big\langle E^{\dagger} (\tau_1) E (\tau_2) E^{\dagger} (\tau_3) E (\tau_4) \big\rangle \bigg) \label{peI}\\
\varrho_{e_i e_j, \text{II}} (t; \Gamma) &= -  \left( - \frac{i}{\hbar} \right)^4 \int^t_{-\infty}\!\!\!\!\!\! d\tau_4 \int^t_{-\infty}\!\!\!\!\!\! d\tau_3 \int^{\tau_3}_{-\infty}\!\!\!\! d\tau_2 \int^{\tau_2}_{-\infty}\!\!\!\!\!\! d\tau_1 \notag \\ 
&\times \bigg( \big\langle V (\tau_4) A_{ij} (t) V (\tau_3) V^{\dagger} (\tau_2) V^{\dagger} (\tau_1) \big\rangle \big\langle E^{\dagger} (\tau_4) E^{\dagger} (\tau_3) E (\tau_2) E (\tau_1) \big\rangle \notag \\
&+ \big\langle V (\tau_1) V (\tau_2) V^{\dagger} (\tau_3) A_{ij} (t) V^{\dagger} (\tau_4) \big\rangle \big\langle E^{\dagger} (\tau_1) E^{\dagger} (\tau_2) E (\tau_3) E (\tau_4) \big\rangle \bigg). \label{peII}
\end{align}
Note that for diagonal density matrix elements, $i.e.$ $e_i = e_j$, we can combine the two terms in both (\ref{peI}) and (\ref{peII}), and recast them as the real part of the first term. To the same order, double-exciton matrix elements are given by diagram III of figure \ref{single2},
\begin{align}
\varrho_{f_i f_j, \text{III}} (t; \Gamma) =  \left( - \frac{i}{\hbar} \right)^4 \int^t_{-\infty}\!\!\!\!\!\! d\tau_4 \int^{\tau_4}_{-\infty}\!\!\!\!\!\! d\tau_3 \int^{t}\!\!\!\! d\tau_2 \int^{\tau_2}\!\!\!\!\!\! d\tau_1 \; &\big\langle V (\tau_3) V (\tau_4) B_{ij} (t) V^{\dagger} (\tau_2) V^{\dagger} (\tau_1) \big\rangle \notag \\
&\times \big\langle E^{\dagger} (\tau_3) E^{\dagger} (\tau_4) E (\tau_2) E (\tau_1) \big\rangle, \label{pf}
\end{align}
\end{widetext}
where $B_{ij} (t) \equiv \vert f_j (t) \rangle \langle f_i (t) \vert$. Eqs. (\ref{peI}), (\ref{peII}) and (\ref{pf}) enable us to evaluate the density matrix for arbitrary states of the radiation field once we specify its four-point field correlation function.

\section{Light sources}
\label{lightsources}
\subsection{The twin photon state}
\begin{figure}[t!]
 \centering
 \includegraphics[width=0.4\textwidth]{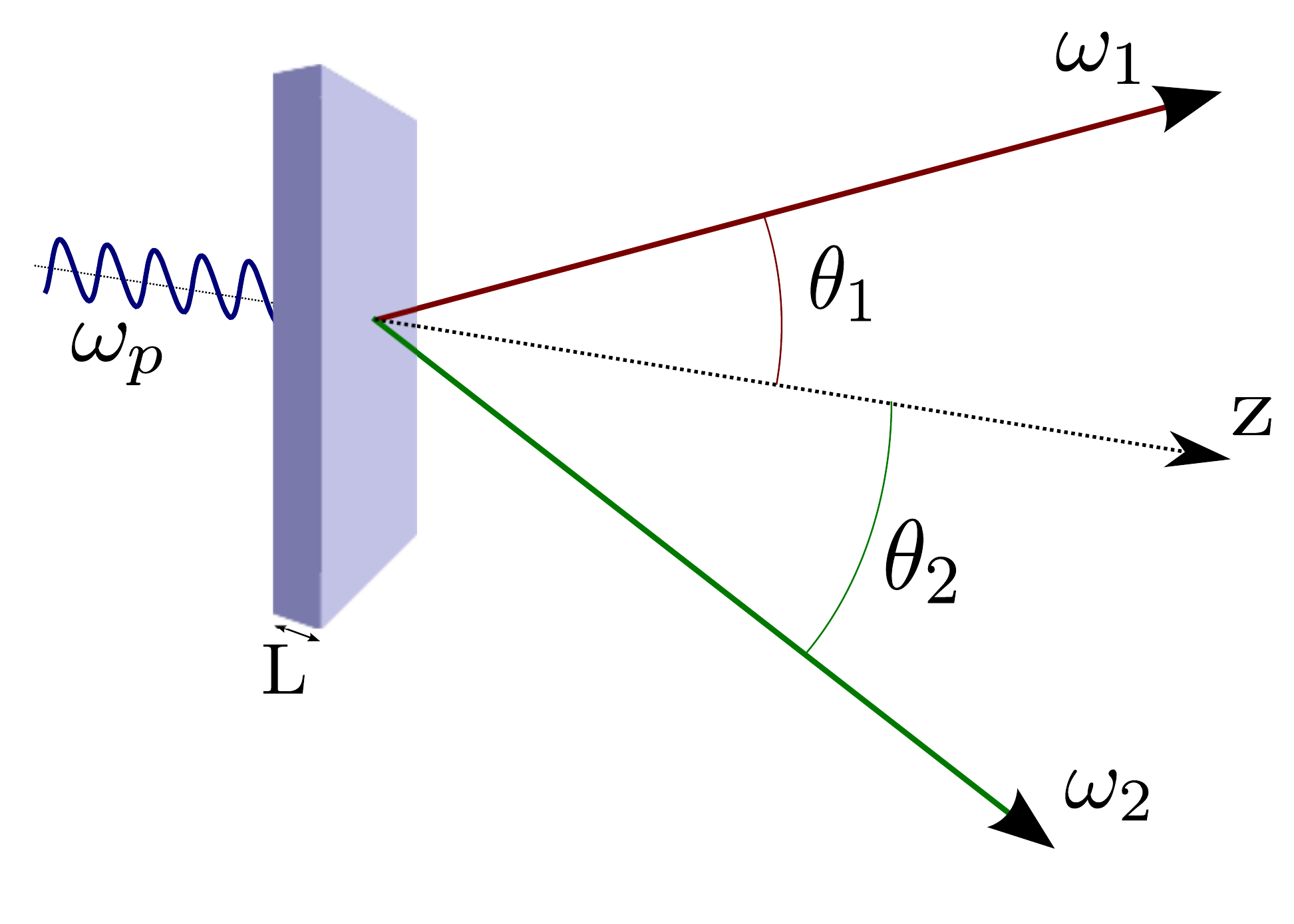}
  \caption{(Color online) Down-conversion setup to produce twin photons: A high-energy pump of frequency $\omega_p$ creates two beams of entangled photons with central frequencies $\omega_1$ and $\omega_2$, such that $\omega_p = \omega_1 + \omega_2$. The central frequencies are determined by the relative angles between the direction of propagation.}
 \label{setup}
\end{figure}
\begin{figure}
 \centering
 \includegraphics[width=.35\textwidth]{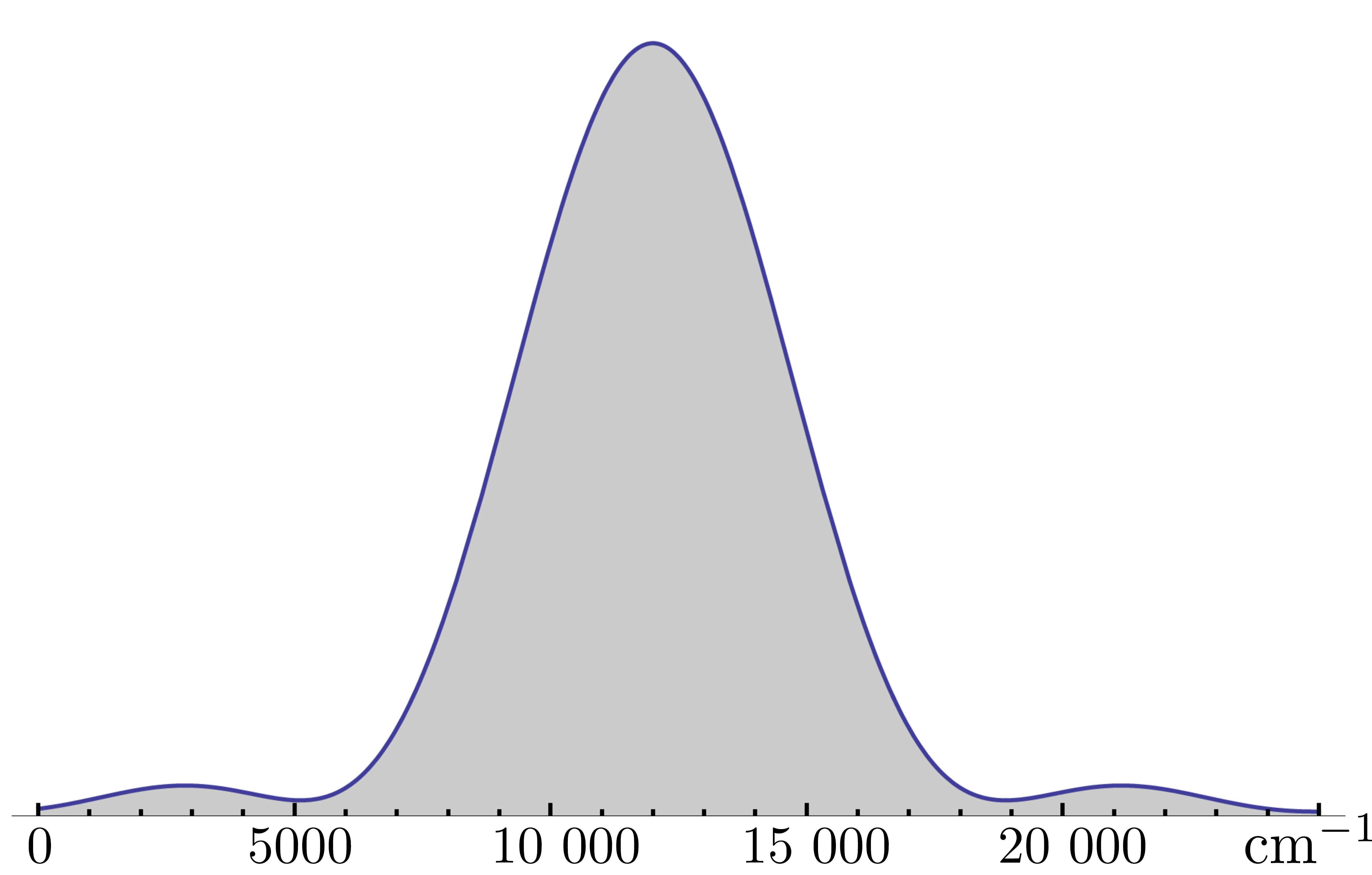}
 \caption{(Color online) Power spectrum of entangled light [eq. (\ref{ntwin})] with $T \; = \; 30 fs$. The central frequencies are $\omega_1 \; = \; 13 000 \; \text{cm}^{-1}$ and $\omega_2 \; = \; 11 000 \; \text{cm}^{-1}$. The two sinc$^2$-functions overlap to form a single broad peak.}
 \label{powerspectrum}
\end{figure}

Extensive research effort has focused on producing and detecting pairs of entangled photons. Numerous schemes have been employed for the generation of entangled photon pairs, including parametric down-conversion \cite{scullyzubairy}, biexciton decay in semiconductors \cite{Tadashi, Stevenson} or four wave mixing in optical fibers \cite{Walmsley1, Camacho}. \\ 
We consider a type of entangled light known as twin photons \cite{Teich2}. It is created by type-II parametric down-conversion with a cw-pump laser of frequency $\omega_p$. When the pump beam is sufficiently weak, it predominantly produces temporally non-overlapping pairs of entangled photons. The state of the field can then be obtained pertubatively in the interaction Hamiltonian of the pump beam with the nonlinear crystal \cite{Teich2}
\begin{widetext}
\begin{align}
\vert \psi \rangle = E_p \int \!\!\! d\vec{k}_1 \int \!\!\! d\vec{k}_2 \;  &\text{sinc}  ( (k_p - k_{1z} - k_{2z}) L / 2 )  e^{i (k_p - k_{1z} - k_{2z}) L / 2 }\notag \\ &\delta (k_{1x} + k_{2 x}) \; \delta (k_{1y} + k_{2 y}) \; \delta ( \omega (\vec{k}_1) + \omega (\vec{k}_2)  - \omega_p ) \vert \vec{k}_1 \vec{k}_2 \rangle, \label{entangledstate}
\end{align}
\end{widetext}
where $E_p$ is the pump field amplitude. The state is not normalized. Here, the length of the crystal along the pump beam is denoted $L$, while it is assumed to be infinitely large in the other two directions. In the setup depicted in figure \ref{setup}, one collects the light in two outgoing directions $\theta_1$ and $\theta_2$ with respect to the optical axis, thus fixing the central frequencies $\omega_1$ and $\omega_2$ of the photon wavepackets in directions 1 and 2. The photon pairs are furthermore correlated in time. This is characterized by the entanglement time $T$, which is determined by the length $L$ along the axis of the pump beam. Therefore, one can vary independently $T$ and two of the three frequencies $\omega_p, \; \omega_1$ and $\omega_2$.\\
For short $T$, eq. (\ref{entangledstate}) leads to the broad power spectrum
\begin{align}
&n_{\text{twin}} (\omega) = \int_{- \infty}^{\infty} dt_1 dt_2 e^{- i \omega (t_1 - t_2)} \langle E^{\dagger} (t_1) E (t_2) \rangle \notag \\
 &= n_0 \vert E_p \vert^2 \left( \text{sinc}^2 \big[ \frac{(\omega - \omega_1)T}{2} \big] + \text{sinc}^2 \big[ \frac{(\omega - \omega_2)T}{2} \big] \right), \label{ntwin}
\end{align}
which is depicted in figure \ref{powerspectrum}.
To evaluate the normally ordered correlation functions in eq. (\ref{peII}), we note that it can be separated into
\begin{align}
&\langle E^{\dagger} (\tau_3) E^{\dagger} (\tau_4) E (\tau_2) E (\tau_1) \rangle = \notag \\ &\langle \psi \vert E^{\dagger} (\tau_3) E^{\dagger} (\tau_4) \vert 0 \rangle \langle 0 \vert E (\tau_2) E (\tau_1) \vert \psi \rangle, \label{factorization}
\end{align}
where \cite{Teich2}
\begin{align}
&\langle 0 \vert E (\tau_2) E (\tau_1) \vert \psi \rangle \notag \\
= &C E_p \; \text{rect} \left( \frac{\tau_2 - \tau_1}{T} \right)   \bigg(e^{-i \omega_1 \tau_1 - i \omega_2 \tau_2} + e^{-i \omega_1 \tau_2 - i \omega_2 \tau_1}\bigg). \label{fourthorder}
\end{align}
Here, we had defined the rectangular function
\begin{align}
\text{rect} (x) = \bigg\{\begin{array}{cc}1, & \text{for} \quad  x  < 1 \\ 0 & \text{else.} \end{array}
\end{align}
The non-classicality of the quantum state produced by this setup has been verified in numerous experiments \cite{Raymer1, twins2, twins3, Silberberg2, Silberberg, Goodson, Goodson2, Teich3}. One important feature that follows from eqs. (\ref{factorization}) and (\ref{fourthorder}) is that the four-point correlation function of the fields scales linearly with the pump intensity $\vert E_p \vert^2$, while classically it scales quadratically \cite{Gould}. This scaling behavior can be utilized in experiments to determine intensity regimes, in which eq. (\ref{entangledstate}) can account for the produced signal. In practice, we expect this linear scaling of fourth-order signals with a crossover to a quadratic classical scaling at higher intensities. The light beam can be considered to be made of pairs of photons. At weak intensities the pairs are temporally well separated, and the process is induced by two photons of the same pair. At higher intensities it becomes statistically more plausible for the two photons to come from different pairs, which are not entangled and the classical scaling is recovered. This crossover has been demonstrated experimentally \cite{ Silberberg, Goodson, Goodson2, Teich3}. For instance, in \cite{Silberberg} the authors detect it at an intensity (of the entangled photon flux) of $1.5 \; \mu W$ in a PPKTP crystal, which was obtained by a pump intensity of less than $2.5 \; W$. The linear scaling of the two-photon absorption was also demonstrated experimentally for organic porphyrin dendrimers \cite{Goodson, Goodson2}. \\
Entangled photon sources are weak, but at the same time weak intensities are required to see the effect of entanglement. Nonlinear spectroscopy and imaging with weak fields should allow to limit the damage to biological samples.

\subsection{Stochastic light with the same power spectrum}
To highlight the effects of entanglement, we consider a classical reference state of the field with the same power spectrum. To that end, we introduce the frequency decomposition of a classical field
\begin{align}
E_i (t) &= \sum_{i} \int d\omega \; A_i (\omega) \; e^{-i \omega t - i \phi (\omega)}, \label{Eclass}
\end{align}
where $A_i$ and $\phi_i$ are real functions representing the amplitude and the phase of the various modes. To reproduce the power spectrum eq. (\ref{ntwin}), we took
\begin{align}
A_1 (\omega) &= A_0 \; \text{sinc} \big[ (\omega - \omega_{1}) T / 2 \big] e^{- i \phi(\omega)}, \label{A_1}\\
A_2 (\omega) &= A_0 \; \text{sinc} \big[(\omega - \omega_2) T/ 2 \big] e^{- i ( \phi(\omega) + \pi / 2)}. \label{A_2}
\end{align}
\begin{figure*}[th!!!]
 \centering
 \includegraphics[width=.8\textwidth]{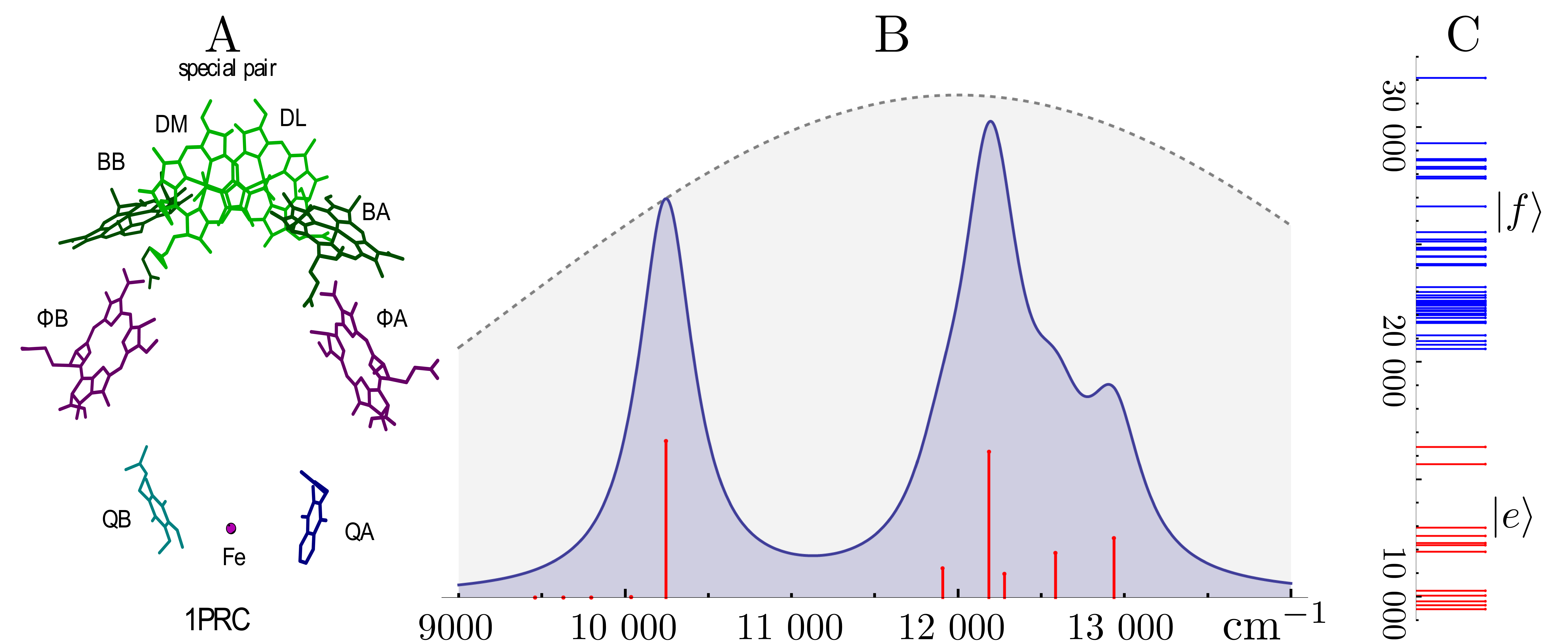}
 \caption{(Color online) A: Arrangement of the six chromophores of the bacterial RC (PDB code: 1PRC) \cite{Deisenhofer}. B: The absorption spectrum of the RC. The lifetime broadening of the excited states is taken to be $200 \; \text{cm}^{-1}$. The gray, dashed line shows  the power spectrum of figure \ref{powerspectrum}, the single exciton states are indicated by red lines, and their heights are proportional to their transition dipole moments $\vert \mu_{ge} \vert^2$. C: Level scheme of the RC. The 12 single-exciton states are marked in red, the 41 double-exciton states in blue.}
 \label{spectrum}
\end{figure*}
With this choice $\phi (\omega)$ does not affect the power spectrum and will be chosen to obtain stationary stochastic light (rather than a pulse). We assume that $\phi (\omega)$ is a random function, so there is no well-defined phase relation between the different frequencies. The quantum mechanical expectation value in correlation functions such as $\langle E^{\dagger} (\tau_3) E^{\dagger} (\tau_4) E (\tau_2) E (\tau_1) \rangle$ then needs to be replaced by an average over the distribution of $\phi (\omega)$:
\begin{align}
&\langle E^{\dagger} (\tau_3) E^{\dagger} (\tau_4) E (\tau_2) E (\tau_1) \rangle\notag \\ 
= &\int d\omega'_1  \int d\omega'_2  \int d\omega'_3  \int d\omega'_4 \; A(\omega'_1) A (\omega'_2) \notag \\
&\times A^{\ast} (\omega'_3) A^{\ast} (\omega'_4) e^{-i (\omega'_2 \tau_2  + \omega'_1 \tau_1)} e^{i (\omega'_4 \tau_4 + \omega'_3 \tau_3)} \notag \\
&\int d \phi(\omega) \; e^{-i (\phi (\omega'_1) + \phi(\omega'_2) - \phi(\omega'_3) - \phi (\omega'_4))}.
\end{align}
Assuming that $\phi$ is uniformly distributed in the interval $[ 0, 2 \pi )$ we find that $E$ is a stationary Gaussian process
\begin{align}
&\langle E^{\dagger} (\tau_3) E^{\dagger} (\tau_4) E (\tau_2) E (\tau_1) \rangle\notag \\ 
= &\langle  E^{\dagger} (\tau_3)E (\tau_2) \rangle \langle E^{\dagger} (\tau_4)  E (\tau_1) \rangle \notag \\
+  &\langle  E^{\dagger} (\tau_4)E (\tau_2) \rangle \langle E^{\dagger} (\tau_3)  E (\tau_1) \rangle.
\end{align}
Classical light only shows correlations on the level of intensities, $i.e.$ $\langle E^{\dagger} E \rangle$. By substitution of eqs. (\ref{A_1}) and (\ref{A_2}) we then obtain
\begin{align}
\langle E^{\dagger} (\tau_3) E^{\dagger} (&\tau_4) E (\tau_2) E (\tau_1) \rangle\notag \\ 
= \mathcal{N} \sum_{\omega_a, \omega_b} \bigg[ &\text{tri} \big( \frac{\tau_1 - \tau_3}{T} \big) \text{tri} \big( \frac{\tau_2 - \tau_4}{T} \big) \notag \\
&\times e^{- i \omega_a (\tau_1 - \tau_3)} e^{- i \omega_b (\tau_2 - \tau_4)} \notag \\
&+ \text{tri} \big( \frac{\tau_1 - \tau_4}{T} \big) \text{tri} \big( \frac{\tau_2 - \tau_3}{T} \big) \notag \\
&\times e^{- i \omega_a (\tau_1 - \tau_4)} e^{- i \omega_b (\tau_2 - \tau_3)} \bigg], \label{classicalcorr}
\end{align}
where the summations run over $\omega_1$ and $\omega_2$, and tri$(x)$ is the triangular function as defined in (\ref{triangular}). This correlation function represents stationary, stochastic light, since it only depends on time differences, just like the one produced by entangled photons with stationary pump [see eq.(\ref{fourthorder})]. However, there is one crucial difference. The quantum mechanical expectation value can be factorized in eq. (\ref{factorization}), and only depends on $\tau_3 - \tau_4$ and $\tau_2 - \tau_1$. The correlation function (\ref{classicalcorr}) depends on $\tau_1 - \tau_3$, $\tau_2 - \tau_4$, $\tau_1 - \tau_4$ and $\tau_2 - \tau_3$. 

\section{Single and double-exciton density matrices in the bacterial reaction center}
\label{system}
We now present simulations of the exciton density matrices produced by entangled and stochastic light with the same spectral density in a model of the bacterial reaction center (RC) of purple bacterium B. viridis \cite{Benjamin}. The primary steps in photosynthesis involve excitation energy transfer towards and charge separation within RCs. The RC of purple bacteria utilizes the high intensity near-IR region of the solar irradiation, and transforms the energy into a chemical potential gradient with near unity efficiency. In bacterial RC the charge separation occurs in the active branch of the protein on the sub-ps to low ps timescale \cite{Zinth2, Zinth3}. The quenching of the excited state special pair population has been demonstrated experimentally in time resolved emission measurements \cite{Hamm1, Tomi}. \\

The electronic Hamiltonian $H_0$ describes the optically bright chromophore excitations and dark charge separated states in the active branch of the RC in a tight-binding formulation. It includes 12 single- and 41 double-exciton states as depicted in figure \ref{spectrum}C. To construct $H_0$, we start with the  X-ray structural data for B. viridis (PDB code: 1PRC) \cite{Deisenhofer}. The Q$_y$ transition dipole moments are placed at the center of bacteriochlorophylls and bacteriopheophytines.  The  excitation energies of BCl's and BP's  are taken from Ref. \cite{Friesner}, F\"orster couplings are calculated in the dipole approximation, except for the special pair  where  a value of 852 cm$^{-1}$ was chosen, allowing to reproduce the CD spectra of B. viridis. The energy of the primary charge separated (CS) state is fixed relative to the lowest special pair exciton state \cite{Parson, Parson2}, which yields a reference point for the additional CS energies of the active branch. The Hamiltonian is given in detail in ref. \cite{Benjamin}.Dephasing effects and incoherent population transport were neglected. To model the absorption spectrum shown in figure \ref{spectrum}, we assume the same lifetime broadening $\gamma \; = \; 200 \; \text{cm}^{-1}$ for all single-exciton states, which agrees reasonably with more detailed simulations \cite{Benjamin}. In figure \ref{spectrum}B, we mark the single-exciton states $e$ by vertical lines, whose length is proportional to their oscillator strength $\vert \mu_{g e} \vert^2$. The spectrum is dominated by six states corresponding to molecular Frenkel excitations of the constituents of the RC. The other six charge transfer states are dark, and cannot be accessed spectroscopically.\\

\begin{figure}[t]
 \centering
 \includegraphics[width=.45\textwidth, angle=-90]{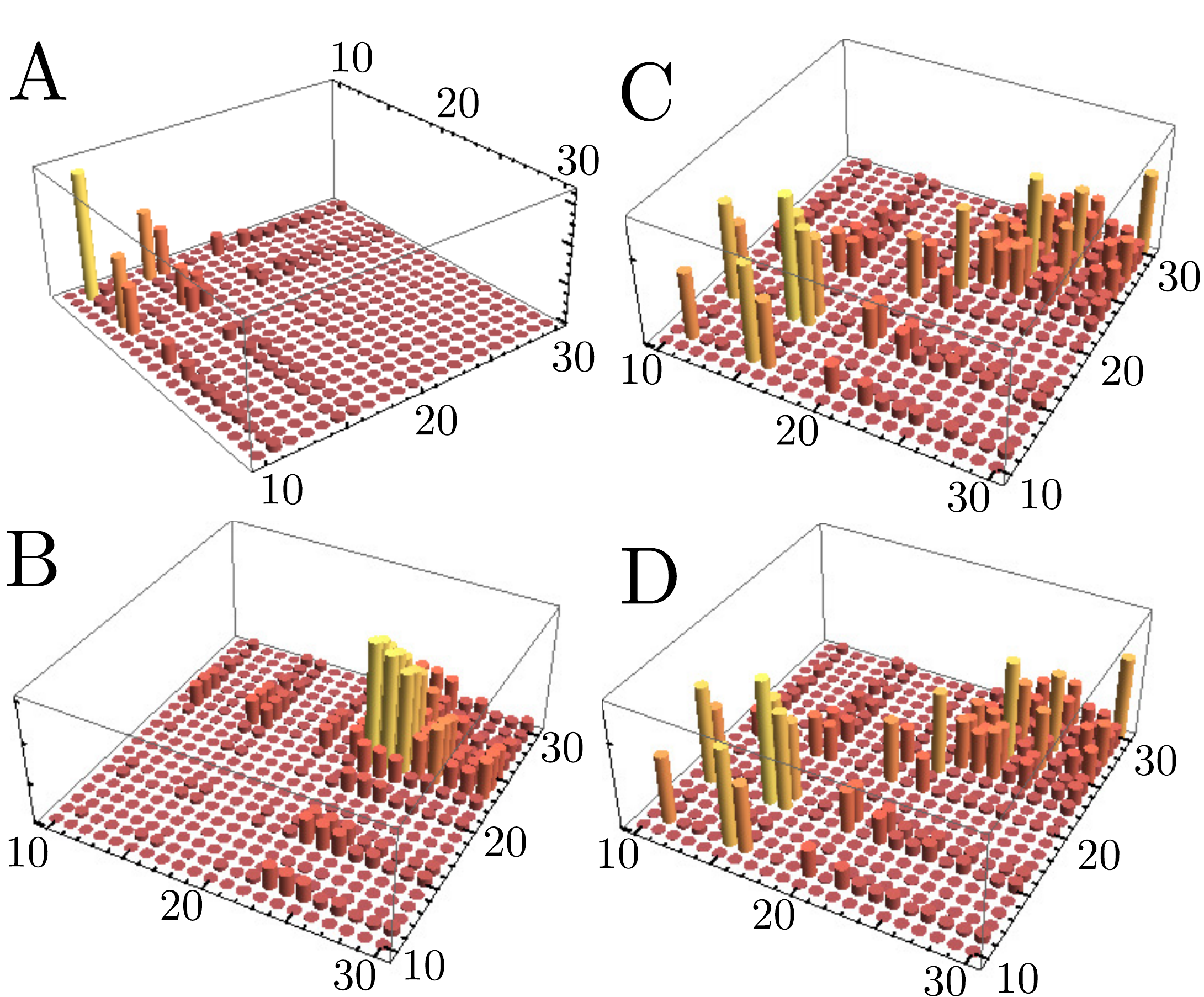}
 \caption{(Color online) Left column: the double-exciton manifold density matrix upon excitation by entangled light with $\omega_p = 22160 \; \text{cm}^{-1}$ (A), and with $24200 \; \text{cm}^{-1}$ (B). Right column: same but for stochastic light. Only the dominant states $f_{10} \dots f_{30}$ are shown.}
 \label{densitymatrix}
\end{figure}
\begin{figure}[t]
 \centering
 \includegraphics[width=.45\textwidth, angle=-90]{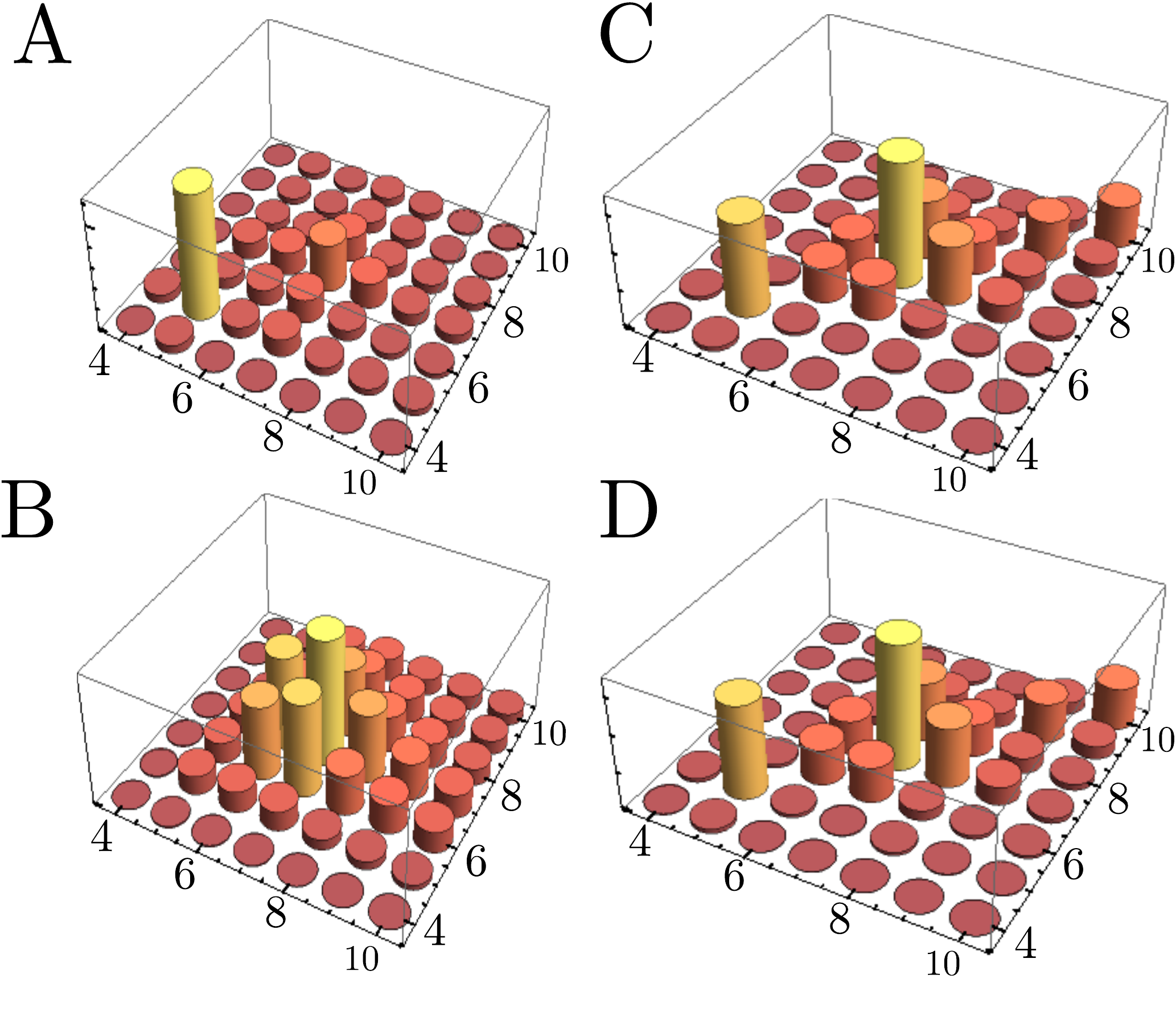}
 \caption{(Color online) Same as figure \ref{densitymatrix}, but for the single-exciton manifold. Onlye dominant states $e_4 \dots e_{10}$ are shown.}
 \label{densitymatrixe}
\end{figure}
The density matrix induced by entangled or stochastic light is block-diagonal in the e- and f-manifolds, since a coherence between the two manifolds can be related to field correlation functions of the kind $\langle E^{\dagger} E^{\dagger} E \rangle$ or $\langle E^{\dagger} E E \rangle$, which vanish for any Fock state or stationary Gaussian process. In figure \ref{densitymatrix} we depict the absolute values of the density matrix elements of the double-exciton states $f_{10}$ to $f_{30}$ for $\omega_p = 22160 \; \text{cm}^{-1}$ and $24200 \; \text{cm}^{-1}$. Entangled light excites a pure double-exciton state, $\sum_f T_{fg} (t) \vert f \rangle$ [see eq. (\ref{Tfgentangled})]. Consequently the purity of the density matrices in figure \ref{densitymatrix}A-B is $\text{tr}_f \{ \varrho_f^2 \} = 1$.  The density matrix induced by stochastic light (figure \ref{densitymatrix}C-D) is not in a pure state, we obtain $\text{tr}_f \{ \varrho_f^2 \} \approx 0.24$ independent of the pump frequency. Furthermore, it is apparent that by tuning the pump frequency, we can select the excitation of certain states. Figure \ref{densitymatrix}A shows strong excitation of the states $f_{11}$, $f_{15}$ and $f_{16}$, while in figure \ref{densitymatrix}B the states between $f_{27}$ and $f_{30}$ are most strongly excited. This selectivity allows to manipulate of the fluorescence signal, as will be shown in the next section. While the density matrices in figures \ref{densitymatrix} A \& B strongly depend the pump frequency, our calculations of the corresponding density matrices produced by stochastic light indicate no such selectivity. The results are plotted in figures \ref{densitymatrix} C \& D. For both pump frequencies ($\omega_1 + \omega_2 = 22160$ and $24200$ cm$^{-1}$), the excitation is distributed among the bright states $f_{11}, f_{15}, f_{22}, f_{25}, f_{27}$ and $f_{30}$.\\
Figure \ref{densitymatrixe} shows the single-exciton manifold density matrices [eq. (\ref{peIIentangled})] for the same pump frequencies $22160$ and $24200$ cm$^{-1}$. Our results qualitatively resemble the ones for the double-exciton manifold. By varying $\omega_p$ we can enhance (figure \ref{densitymatrixe}A) or suppress (figure \ref{densitymatrixe}B) the excitation of $e_5$ (and, vice versa, of $e_6$ and $e_7$), while stochastic light always creates very similar density matrices. For both frequencies, the population is mainly distributed between the states $e_5$ and $e_7$.
\begin{figure}[h]
\centering
\includegraphics[width=.225\textwidth]{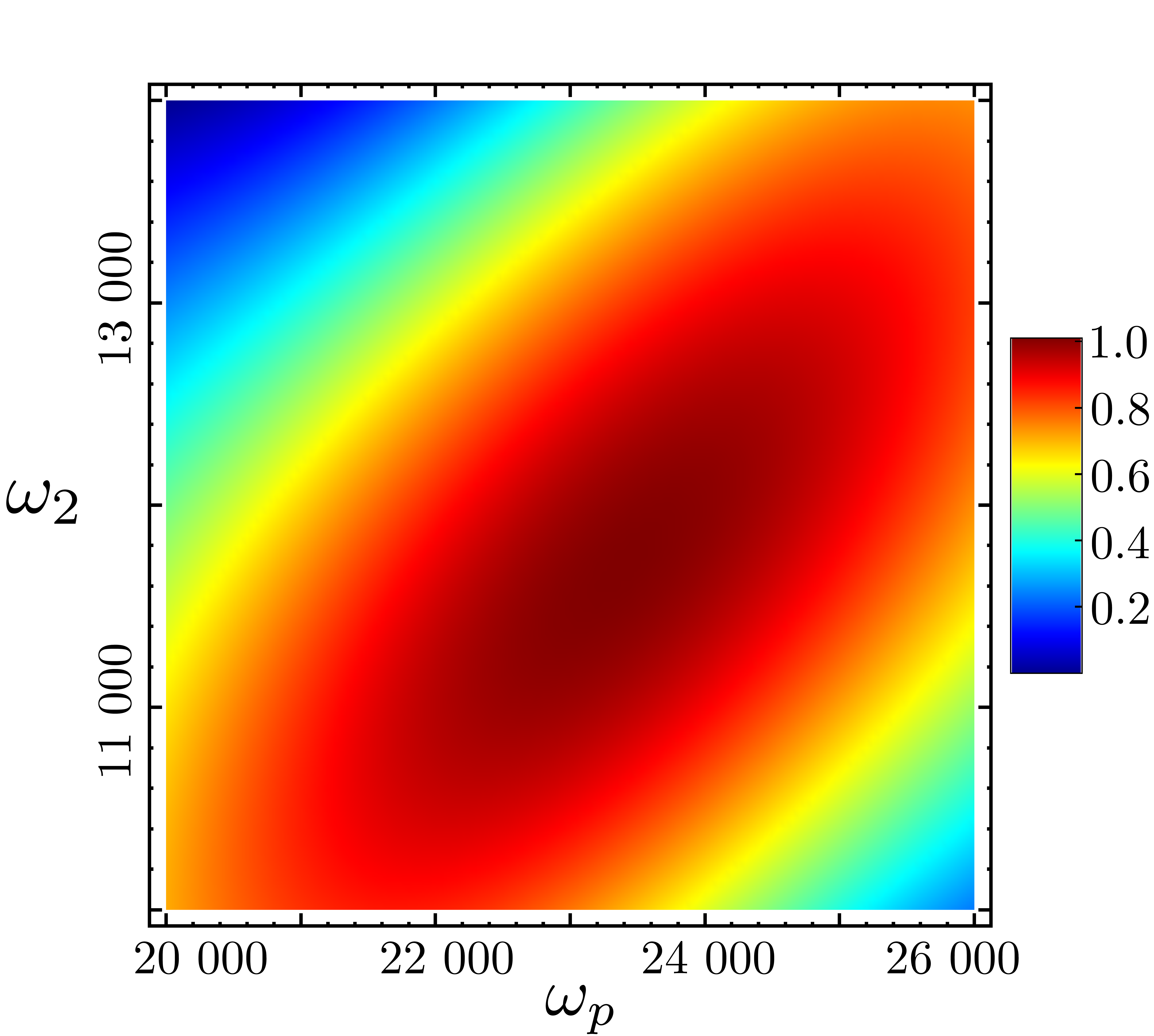}
\caption{(Color online) Second-order contribution to the total single-exciton manifold population $\sum_e \varrho_{ee} (t)$ , eq. (\ref{S2a}), with $T \; = \; 30 \; fs$.}
\label{p'e2D}
\end{figure}
\begin{figure}[t]
 \centering
 \includegraphics[width=.45\textwidth]{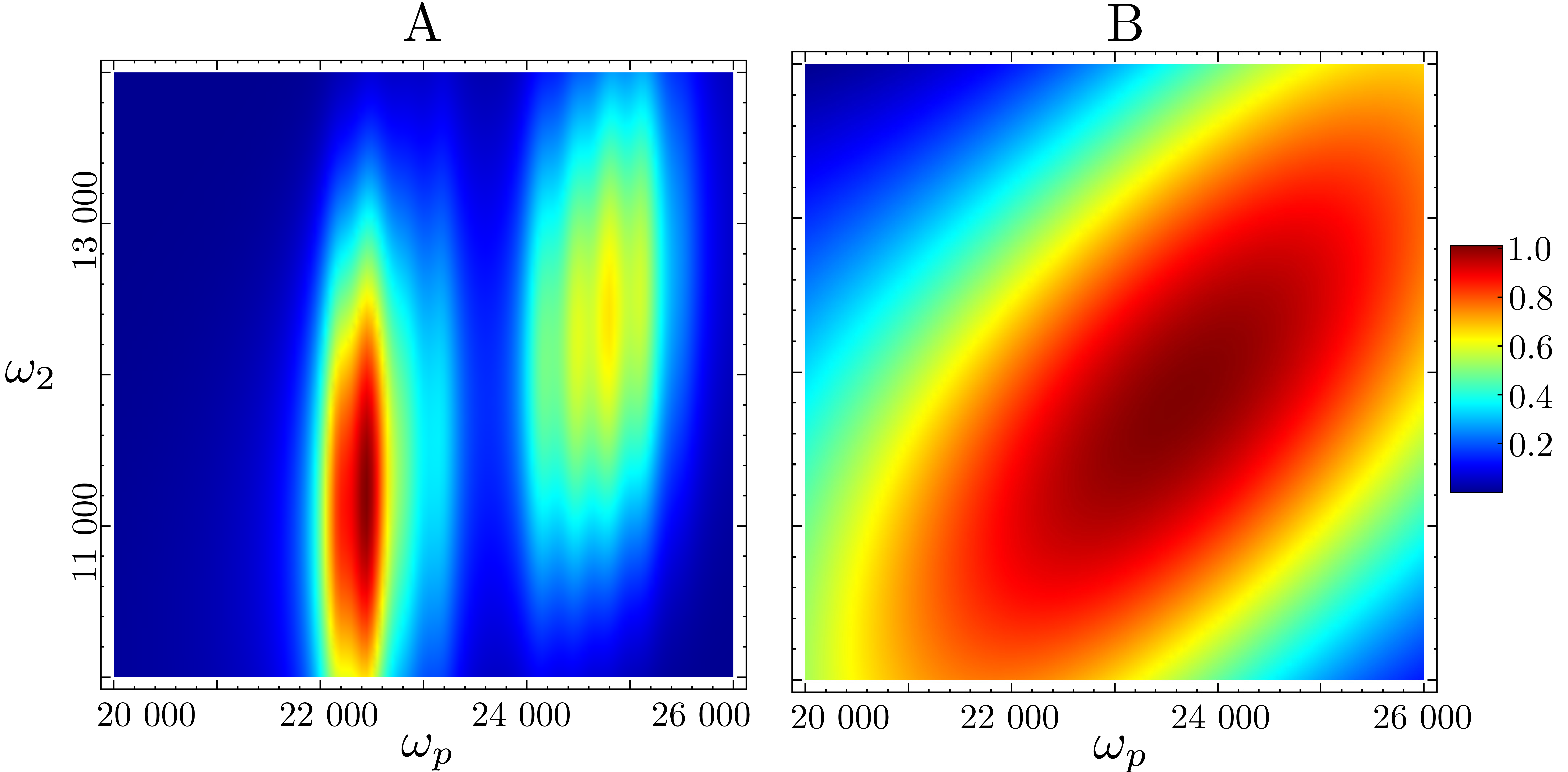}
 \caption{(Color online) Total population of the double-exciton manifold $\sum_f \varrho_{f f}$ [eq. (\ref{pf})] plotted vs. $\omega_2$ and $\omega_p$. A: excitation by entangled light with $T \; = \; 30 \; fs$. B: Excitation by stochastic light.}
 \label{manifolds}
\end{figure}

The total population of the single-exciton manifold $\sum_e \varrho'_{ee}$ calculated to second order in the matter-field interaction [eq. (\ref{p'eentangled})]  is displayed in figure \ref{p'e2D} vs. the frequencies $\omega_p$ and $\omega_2$. Due to the broadband nature of the beams, none of the peaks of the absorption spectrum in figure \ref{spectrum}B can be resolved. This population only depends on the spectral density, is identical for stochastic and entangled light, and is not susceptible for the non-classical features of entangled light.\\
The total double-exciton population $\sum_f \varrho_{f f}$ created by eq. (\ref{Tfgentangled}) is plotted in figure \ref{manifolds}A. The distribution of the single-exciton states to fourth order (see diagram II in figure \ref{single2}) looks almost identical, since eq. (\ref{peII}) includes an excitation to the double-exciton manifold, and is not shown. It depends only weakly on $\omega_2$, reflecting the energy uncertainty due to the time entanglement. Thus, the central frequency of each beam is not important, and in the following we set $\omega_2 = 11 000 \; \text{cm}^{-1}$. Two broad resonance bands, at $22 000 - 23 000 \; \text{cm}^{-1}$ (band $a$), and at $23 000 - 24 000 \; \text{cm}^{-1}$ (band $b$), dominate the distribution. From our previous discussion of the density matrix, we anticipate that states $f_{11}, f_{15}$ and $f_{16}$ dominate band $a$, whereas the states $f_{23} \; - \; f_{27}$ seem to make up band $b$ (see figure \ref{densitymatrix}). The signal induced by stochastic light (figure \ref{manifolds}B) shows no such structure. Due to the broad bandwidth, the level structure of the RC cannot be resolved at all. \\
\begin{figure}
 \centering
 \includegraphics[width=.45\textwidth]{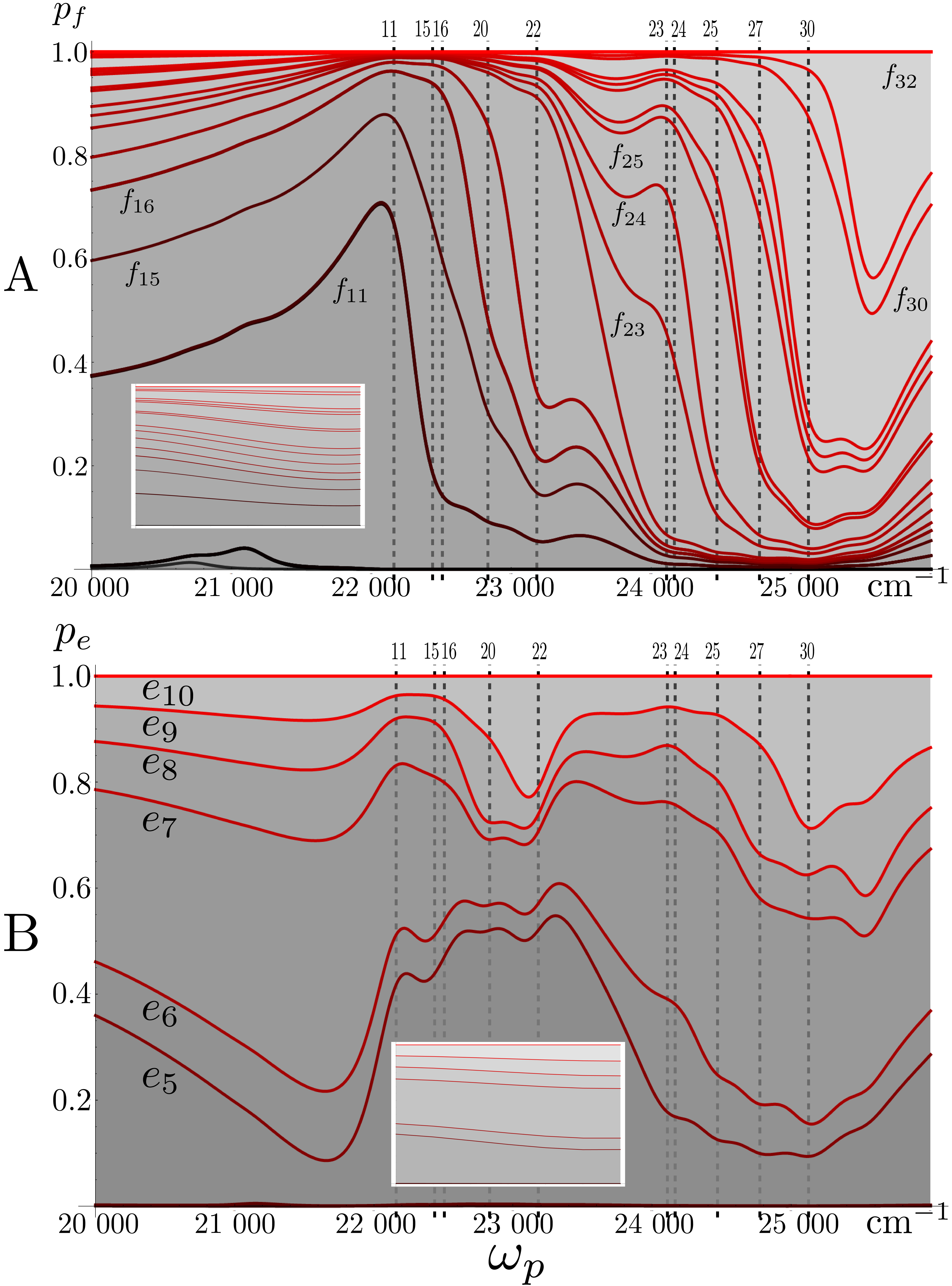}
 \caption{(Color online) A: contributions of the various f-states to the total population for $\omega_2 \; = \; 11 000 \; cm^{-1}$ with $T \; = \; 30 \; fs$ upon excitation by entangled light. Insets: Excitation by stochastic light. B: same for the single-exciton states}
 \label{pfsingle}
\end{figure}
To get a more detailed picture of this state distribution, we depict the contribution of various double-exciton state populations to the distribution [eq. (\ref{Tfgentangled})] in figure \ref{pfsingle}A. The distributions are normalized at each $\omega_p$. States $f_{11}$, $f_{15}$ and $f_{16}$ dominate band $a$, whereas band $b$ is dominated by states $f_{23}$, $f_{24}$ and $f_{25}$. At higher pump frequency, states $f_{27}$, $f_{30}$ and $f_{32}$ are most pronounced. In general, the regions of leading contribution group around the states' energies, but the distributions may be asymmetric because of the presence of other states. We next turn to the distribution of single-exciton states (figure \ref{pfsingle}B). Since the single excitons are obtained by emission from a two-exciton state, the distribution closely resembles the double-exciton distribution in figure \ref{pfsingle}A. For instance, when $\omega_ p \sim 22 200 \; \text{cm}^{-1}$, the double-exciton state $f_{11}$ is on resonance. It decays primarily into single-exciton state $e_5$, and accordingly the contribution of this state increases  around the same value of the pump frequency. The same holds for, e.g., the double-exciton state $f_{30}$, and the single-exciton state $e_{10}$. However, in most cases the different contributions overlap, and the relation between the single-exciton and the double-exciton states is more complex.

\section{Two-photon-induced fluorescence}
\begin{figure}[ht!]
 \centering
 \includegraphics[width=0.25\textwidth, angle=-90]{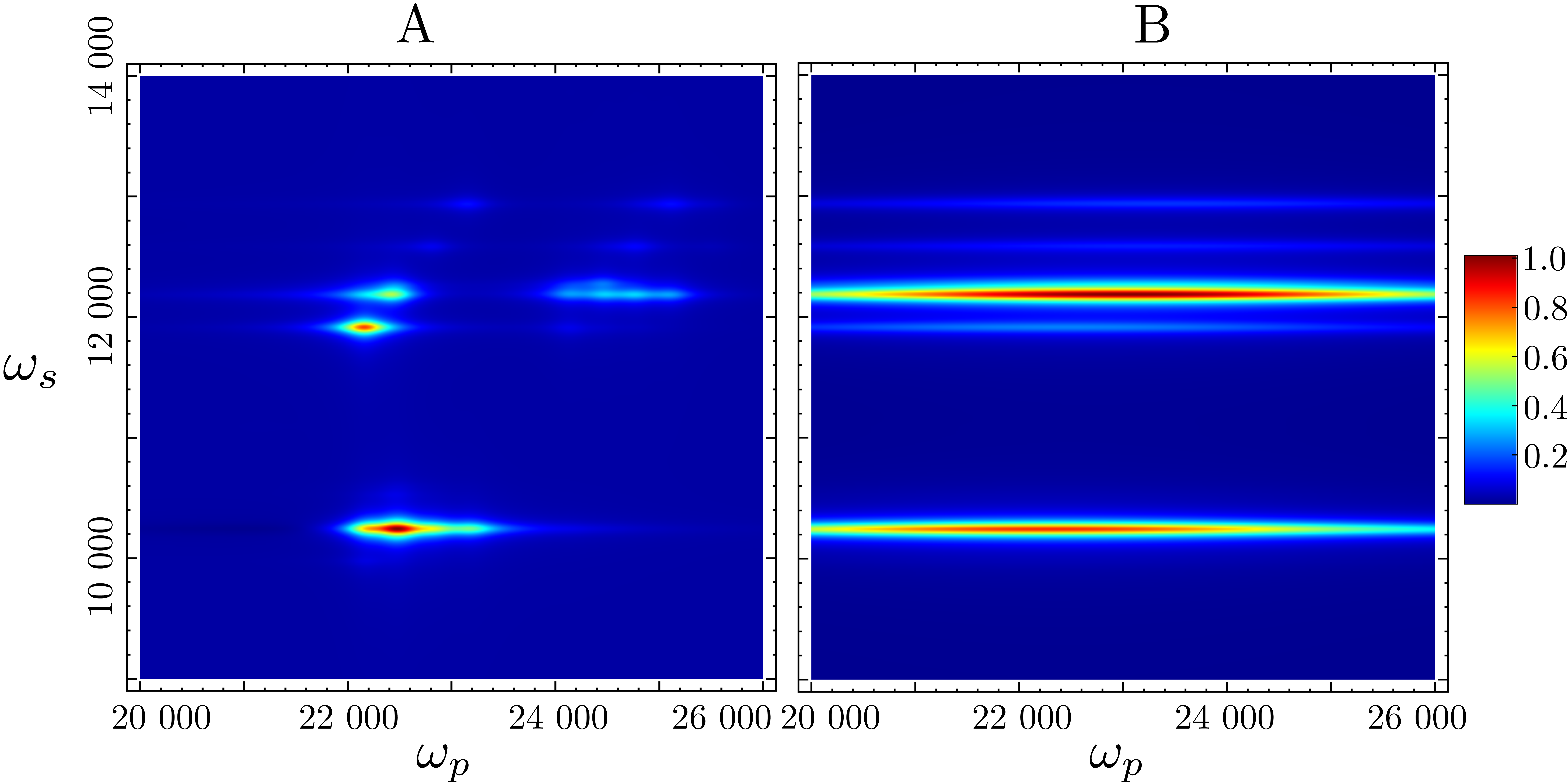}
 \caption{(Color online) A: 2D spectrum of the fluorescence signal of entangled photons with entanglement time $T \; = \; 30 fs$. B: The corresponding spectrum with classical stochastic light. We set the frequency $\omega_2 = 11 \; 000 \; cm^{-1}$. Two-exciton peaks are indicated by vertical black lines, and single-exciton peaks by horizontal lines.}
 \label{2D3fs}
\end{figure}
\begin{figure}[t!]
 \centering
 \includegraphics[width=.45\textwidth]{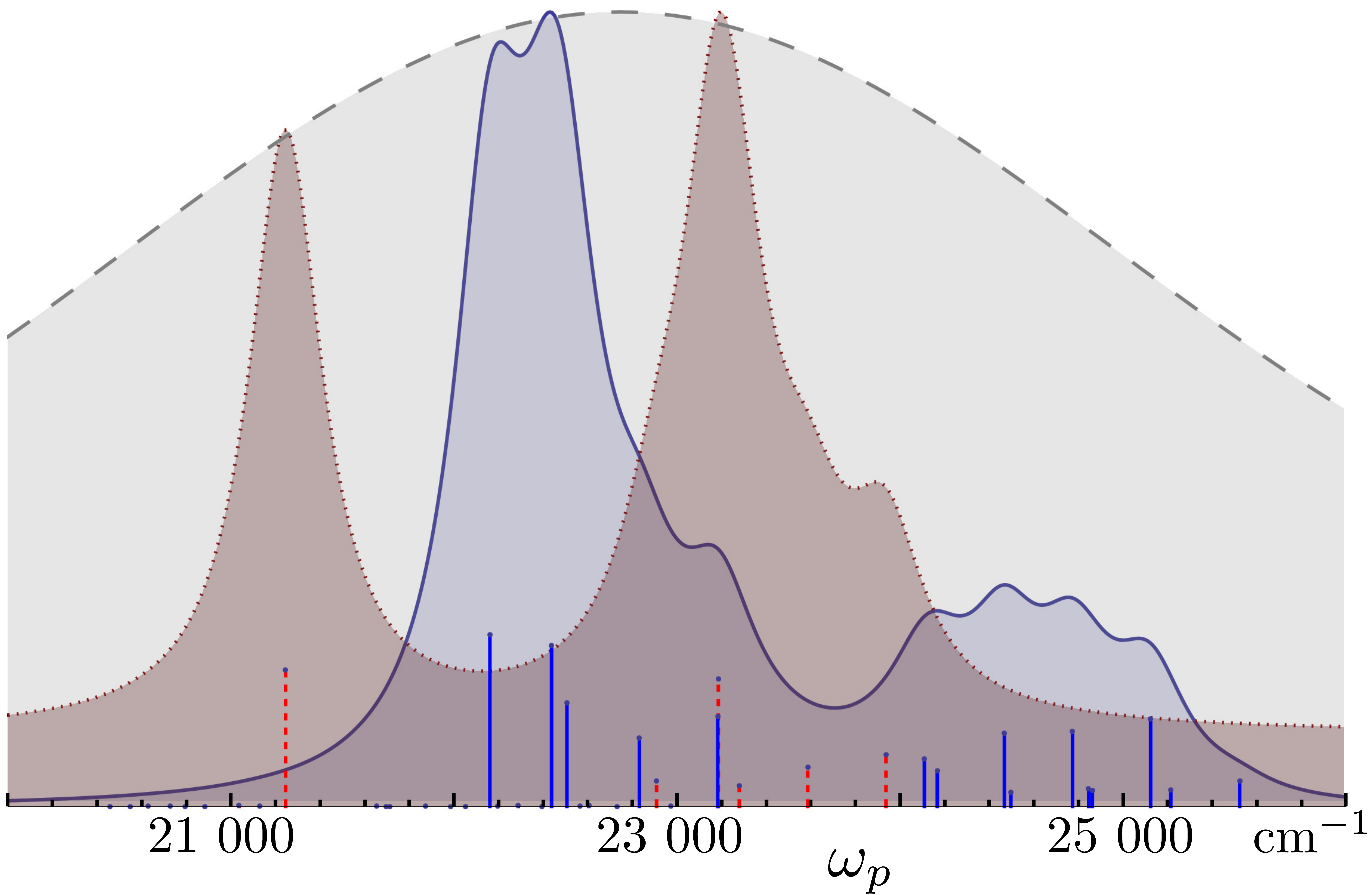}
 \caption{(Color online) The fluorescence action spectrum (integrated over the emission frequency $\omega_s$) vs. the pump frequency $\omega_p$: signal created by entangled photons (solid, blue line), the stochastic light as (dashed, gray line), and signal created by two monochromatic beams at $\omega_1$ and $\omega_2$ (dotted, red line).\newline
The double-exciton energies are plotted as solid, blue lines, and their heights are given by $\vert\sum_e \mu_{ge} \mu_{ef}\vert^2$. The single-exciton energies are indicated as dashed, red lines, with the height proportional to $\vert \mu_{ge} \vert^2$.}
 \label{wsintegrated}
\end{figure}

Having discussed our simulations of the excited state distribution created by entangled light, we now turn to the question whether it may be possible to detect fingerprints of these distributions in experiments. To that end, we calculate the dispersed fluorescence signal. This involves transitions between double- and single-exciton states as well as transitions between single-exciton states and the ground state, $i.e.$
\begin{align}
S_f (\omega_s; \Gamma) &=  \sum_{e,f} \vert \mu_{fe} \vert^2  p_f (t; \Gamma) \delta (\omega_f - \omega_e - \omega_s), \label{newsignal} \\
S_e (\omega_s; \Gamma) &= \sum_e \vert \mu_{eg} \vert^2 p_e (t; \Gamma) \delta (\omega_e - \omega_s). \label{signal2}
\end{align}
Here, $\omega_s$ is the emitted fluorescence frequency, and $p_f$ ($p_e$) the population of state f (e) given by eqs. (\ref{pf}) and (\ref{peII}), respectively. More elaborate detection (gated time and frequency, photon statistics) is possible \cite{Konstantin2}, but will not be considered here. \\
In figure \ref{2D3fs}, we display this signal vs. the pump frequency $\omega_p$ and the emission frequency $\omega_s$. The delta functions in (\ref{newsignal}) and (\ref{signal2}) were replaced by a Lorentzian, which means the width of the peaks in horizontal direction of figure \ref{2D3fs} is instrumental. In analogy to the populations shown in figure \ref{manifolds}, the fluorescence simulation with entangled photons (\ref{2D3fs}A) shows two distinct resonances along the $\omega_p$-axis pertaining to bands $a$ and $b$, respectively (see also figure \ref{wsintegrated}). Along $\omega_s$ (horizontal axis), it contains two contributions, one around $\omega_s \sim 10 000 \; \text{cm}^{-1}$, and a higher-energy part between $12 000$ and $13 000 \; \text{cm}^{-1}$, reminiscent of the absorption spectrum. This pronounced structure can be exploited to enhance or suppress certain features, as will be demonstrated in the following. The simulation of the stochastic signal in figure \ref{2D3fs}B shows no structure along the $\omega_p$-axis, and consequently does not allow a manipulation of the fluorescence signal.
\begin{figure}[t!]
 \centering
 \includegraphics[width=0.47\textwidth]{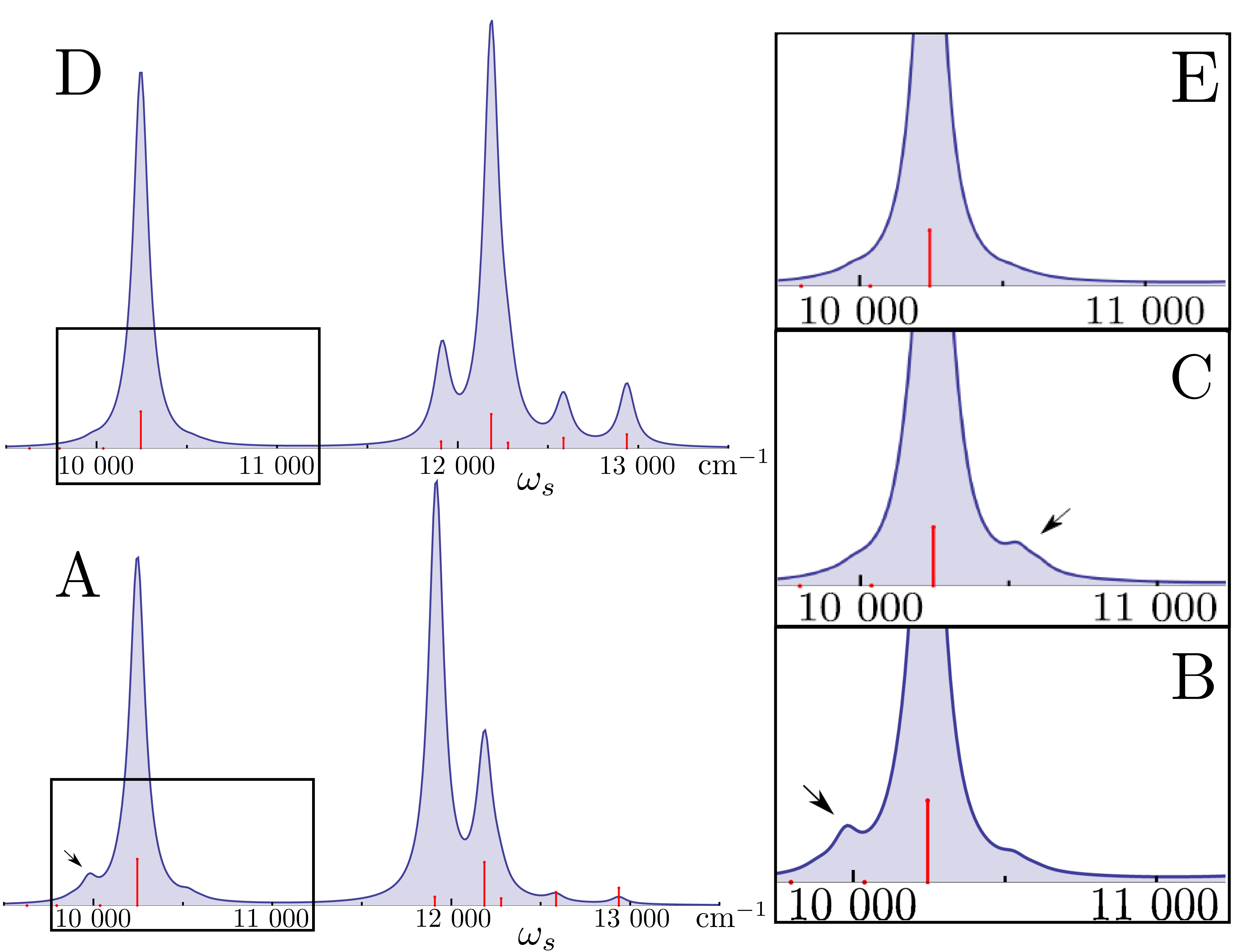}
 \caption{(Color online) A: The dispersed fluorescence signal induced by entangled photons with entanglement time $T \; = \; 30 fs$ and $\omega_p \; = \; 22160 \; \text{cm}^{-1}$, B: enlarged region by the black box. C: Enlarged region for $\omega_p \; = \; 22500 \; \text{cm}^{-1}$. D \& E: same as A \& B, but for stochastic light.}
 \label{wp22160}
\end{figure}
The dispersed fluorescence simulation for $\omega_p \; = \; 22160 \; \text{cm}^{-1}$ is shown in figure \ref{wp22160}A. Most of the peaks can be directly attributed to the energy of single-exciton states, the transition thus correspond to either fluorescence from the single-exciton manifold, or from double-exciton states, whose energies closely match the sum of two single-exciton states. However, the simulations also show side peaks of the resonance at $10 245 \; \text{cm}^{-1}$, which cannot be assigned to single-exciton energies. The strongest of these peaks is highlighted by the black arrow in the inset figure \ref{wp22160}B. It corresponds to the $f_{11} \rightarrow e_{7}$ transition. Since state $f_{11}$ is most strongly excited at the given pump frequency (see figure \ref{pfsingle}A), this peak is most pronounced. At pump frequency $22 500 \; \text{cm}^{-1}$ resonantly excited states $f_{15}$ and $f_{16}$ show different transitions into the single-exciton state $e_6$ (see figure \ref{wp22160}C). The simulation with stochastic light shown in figures \ref{wp22160}D and E cannot resolve any of these sidepeaks due to its broadband nature.

\section{Conclusions}
We had derived expressions for the single- and double-exciton density matrices of quantum systems interacting with arbitrary light sources perturbatively in the field. These were used to simulate the excitation of matter via entangled twin photons, and stochastic light with the same power spectrum. Applications to the reaction center of purple bacteria show that the populations strongly depend on the nature of the light, and we indicated how these properties could be observed in the frequency-resolved fluorescence measurements. The non-classical spectral profile of entangled light allows us to target specific double-exciton states, and to explore all the excitation pathways to this state in a single shot. This shows up perspiciously in the fluorescence action spectrum in figure \ref{wsintegrated}. The spectrum reveals the level structure of the double-exciton manifold, whereas the stochastic light cannot resolve this structure. For completeness, we also plot the action spectrum created by two cw laser beams with frequencies $\omega_1$ and $\omega_2$ at the peaks of the spectral density. Even though the narrow bandwidth of these beams allows for a good $\omega_p$-resolution, this signal is most pronounced, when one of the laser beams is resonant with a single-exciton state, and thus closely resembles the linear spectrum in figure \ref{spectrum}.\\
The role of entanglement and coherent transport in the efficiency of exciton transport in photosynthetic complexes has been subject of intense debate \cite{ Engel, Scholes, Aspuru-Guzik, Plenio1, Plenio2, Torsten, Kauffmann2, Kauffmann3}. Using our formalism, one could also discuss the entanglement of quasiparticles in the double-exciton manifold \cite{Shaul3, Mintert}. This will require to decompose double-exciton states into single excitons, which goes beyond the scope of the present Paper.\\
Other possible extensions are to excitation by pulsed light, and to include time- and frequency gating in order to obtain control over the time and frequency resolution of the detection. The time resolution could allow us to monitor the dynamics of exciton transport and charge separation after interaction with entangled light in an experimentally feasible way. 

\section{Acknowledgements}
We gratefully acknowledge the support of the National Science Foundation through Grant No. CHE-1058791, the Chemical Sciences, Geosciences and
Biosciences Division, Office of Basic Energy Sciences, Office of Science, US Department of Energy, and the Defense Advanced Research Projects
Agency (DARPA), grant UOFT-49606. B. P. F. gratefully acknowledges support from the Alexander-von-Humboldt Foundation through the Feodor-Lynen program.

\appendix
\section{Excitations induced by twin photons}
\label{appendixa}
Using eq. (\ref{entangledstate}), the field correlation function of the leading-order contribution, given by eq. (\ref{S2a}), yields 
\begin{align}
&\langle E^{\dagger} (\tau_2) E (\tau_1) \rangle \notag \\ = &C \; \text{tri} \left( \frac{\tau_2 - \tau_1}{T} \right) \left( e^{- i \omega_1 (\tau_1 - \tau_2)} + e^{- i \omega_2 (\tau_1 - \tau_2)} \right),
\end{align}
where we have defined the triangular function
\begin{align}
\text{tri} (x) &= \bigg\{\begin{array}{c}1 - \vert x \vert, \qquad \qquad \text{for} \quad \vert x \vert < 1 \\ \!\!\!\!\!\!\!\!\!\!\!\!\!\!\! 0 \qquad \qquad \qquad \text{else.} \end{array} \label{triangular}
\end{align}
Expanding the matter correlation function into a sum-over-states expression, we obtain
\begin{align}
\varrho'_{e_i, e_j} (t; \Gamma) &= \!\! \frac{i \mu_{g e_i} \mu_{g e_j}}{4 \hbar^4 (\omega_{e_i e_j} + 2 i \gamma_e)} \bigg( \text{sinc} \left( (\omega_1 - \omega_{e_i g} + i \gamma_{e_i}) T / 2\right) \notag \\
&+ \text{sinc} \left( (\omega_2 - \omega_{e_j g} - i \gamma_{e_j}) T / 2\right) \bigg), \label{p'eentangled}
\end{align}
where we dropped the constant factor C. The matrix element of the dipole operator connecting the states g and e is denoted $\mu_{ge}$, the energy difference is given by $\omega_{eg}$, and the lifetime broadening is denoted by $\gamma_e$. We can assume that $\gamma_{e_i} \approx \gamma_{e_j}$. Since the entanglement time $T$ is short with relative to other timescales in the system ($T \sim 10^{-3} \; \text{cm}^{-1}$), this contribution to the single-exciton states depends weakly on the pump frequency $\omega_p$.\\
The field correlation function of eq.(\ref{peI}) can be recast as a normally ordered term plus a commutator term, which we neglect. Using eq. (\ref{fourthorder}) and dropping the constant factor $C$, we obtain for the populations
\begin{align}
\varrho_{e, e \text{I}} (t; \Gamma) = &\Re \bigg\{ \sum_{e ,e'} T \frac{ \mu_{g e}^2 \mu_{g e'}^2}{\gamma_e \hbar^4} \frac{e^{i (\omega_1 - \omega_{e' g} + i \gamma_{e'}) T} - 1}{\omega_1 - \omega_{e' g} + i \gamma_{e'}}  \notag \\
&\frac{e^{ i (\omega_2 - \omega_{e g} + i \gamma_e) T} - 1}{\omega_2 - \omega_{e g} + i \gamma_e} \bigg\}.
\end{align}
This contribution is proportional to the entanglement time $T$, reflecting the fact that in (\ref{peI}) the time pairs ($\tau_1$, $\tau_3$) as well as ($\tau_2$, $\tau_4$) are correlated. This means that the entire process has to happen within the entanglement time $T$. Since this time is very short with respect to other timescales in the system, $\varrho_{e, e I}$ can be neglected. The only fourth-order contribution to the single-exciton manifold is thus given by (\ref{peII}). In a sum-over-states expression, it reads
\begin{widetext}
\begin{align}
\varrho_{e_i, e_j \text{II}} (t; \Gamma) =  &\sum_{e', f} \sum_{a, b, c, d} \frac{\mu_{g e'} \mu_{e' f}}{\hbar^4} \bigg[ \frac{\mu_{f e_i} \mu_{g e_j}}{\omega_1 + \omega_2 - \omega_{fg} + i \gamma_f} \frac{e^{i (\omega_a - \omega_{e_i g} + i \gamma_{e_i}) T} - 1}{\omega_a - \omega_{e_i g} + i \gamma_{e_i}} \frac{e^{i (\omega_b - \omega_{e' g} + i \gamma_{e'}) T} - 1}{\omega_1 - \omega_{e' g} + i \gamma_{e'}} \notag \\
&- \frac{\mu_{f e_j} \mu_{g e_i}}{\omega_1 + \omega_2 - \omega_{fg} -  i \gamma_f}
 \frac{e^{-i (\omega_c - \omega_{e_j g} - i \gamma_{e_j}) T} - 1}{\omega_c - \omega_{e_j g} - i \gamma_{e_j}} \frac{e^{-i (\omega_d - \omega_{e' g} - i \gamma_{e'}) T} - 1}{\omega_d - \omega_{e' g} - i \gamma_{e'}}
\bigg], \label{peIIentangled}
\end{align}
\end{widetext}
where the summations a, b, c and d runs over $\omega_1$ and $\omega_2$. It is apparent that the single-exciton peaks are multiplied by phase factors of the kind $e^{ i (\omega_1 - \omega_{e g} + i \gamma_{e}) T} - 1$, while the two-exciton peak is not. Additionally, taking into account
\begin{align}
&\frac{e^{i (\omega - \omega_{eg} + i \gamma_e) T} - 1}{\omega- \omega_{eg} + i \gamma_e} \approx\notag \\ 
&2 i T e^{i (\omega - \omega_{eg})T / 2} \text{sinc} \left( (\omega - \omega_{eg}) T / 2 \right),
\end{align}
the single-exciton peaks have a very similar structure as (\ref{p'eentangled}), and depend weakly on $\omega_p$. Thus, $\varrho_{e_i, e_j} (t)$ is dominated by the Lorentzian $1 / (\omega_1 + \omega_2 - \omega_{fg} + i \gamma_f)$, and this in turn means that spectroscopy with entangled photons can be used to probe the two-exciton manifold of aggregates.\\
We further note that eq. (\ref{pf}) can be recast as the product of transition amplitudes $T_{fg} (t)$ given by
\begin{widetext}
\begin{align}
T_{fg} (t) &= \left( - \frac{i}{\hbar} \right)^2 \int^t_{-\infty} d\tau \int^{\tau}_{-\infty} d\tau' \; \langle f(t) \vert V^{\dagger} (\tau) V^{\dagger} (\tau') \vert g \rangle \langle 0 \vert E (\tau) E (\tau') \vert  \psi \rangle \\
&= \frac{1}{\hbar^2} \sum_e \frac{\mu_{ge} \mu_{ef}}{\omega_1 + \omega_2 - \omega_{fg} + i \gamma_f} \left( \frac{e^{i (\omega_1 - \omega_{e g} + i \gamma_{e}) T} - 1}{\omega_1 - \omega_{e g} + i \gamma_{e}} + \frac{e^{i (\omega_2 - \omega_{e g} + i \gamma_{e}) T} - 1}{\omega_2 - \omega_{e g} + i \gamma_{e}} \right) e^{- i (\omega_1+ \omega_2)t}.  \label{Tfgentangled}
\end{align}
\end{widetext}
Clearly, the single-exciton resonances are suppressed, and the detuning between $\omega_p$ and the two-exciton manifold completely determines $\varrho_{f_i, f_j} (t; \Gamma)$.

\section{Electronic excitations with stochastic light}
Due to the different intensity scaling of classical signals, the main contribution to the signal will arise from the leading-order contribution in eq. (\ref{S2a}). Since the stochastic light reproduces the spectral density of entangled light, we obtain the same result as for entangled light [see eq. (\ref{p'eentangled})]. Additionally, one can filter out this contribution, and calculate (\ref{peII}) and (\ref{pf}) to study the electronic excitations induced by higher-order interaction with stochastic light.\\
Using eqs. (\ref{peI}), (\ref{pf}) and (\ref{classicalcorr}), the fourth-order contributions to the single-exciton manifold yield
\begin{widetext}
\begin{align}
\varrho_{e_i, e_j \text{II}} (t; \Gamma) &= \frac{1}{8 \hbar^4} \sum_{a, b} \sum_{e'} \mu_{g e'} \mu_{e' f} \text{sinc}^2 \left( (\omega_a - \omega_{e' g}) T / 2 \right) \text{sinc}^2 \left( (\omega_b - \omega_{f e'}) T / 2 \right) \notag \\
&\times \bigg( \mu_{g e_j} \mu_{f e_i} \left( \frac{1}{(\omega_{fg} - \omega_{e' g} - \omega_{e_i g} + i \gamma_{f})(\omega_{fg} - \omega_{e' g} - \omega_{e_j g} - i \gamma_{f})} - 
\frac{1}{(\omega_{e' e_j} - 2 i \gamma_e)(\omega_{e_j e_i} - 2 i \gamma_e)} \right) \notag \\
&+ \mu_{g e_i} \mu_{f e_j}\left( \frac{1}{(\omega_{fg} - \omega_{e' g} - \omega_{e_i g} - i \gamma_{f})(\omega_{fg} - \omega_{e' g} - \omega_{e_j g} + i \gamma_{f})} - 
\frac{1}{(\omega_{e' e_i} - 2 i \gamma_e)(\omega_{e_i e_j} - 2 i \gamma_e)} \right)\bigg), \label{peIIstochastic}\\
\varrho_{f_i, f_j} (t; \Gamma) &= \frac{1}{16 \hbar^4} \sum_{a, b} \sum_{e, e'} \mu_{ge} \mu_{e f_i} \mu_{g e'} \mu_{e' f_j} \text{sinc}^2 \left( (\omega_a - \omega_{eg}) T / 2 \right) \text{sinc}^2 \left( (\omega_b - \omega_{f_j e}) T / 2 \right) \notag \\
&\times  \bigg( \frac{1}{(\omega_{f_j g} - \omega_{e g} - \omega_{e' g} - i \gamma_f)(\omega_{f_i g} -\omega_{e g} - \omega_{e' g} + i \gamma_f)} - \frac{1}{(\omega_{e' e} + 2 i \gamma_e)(\omega_{f_i f_j} + 2 i \gamma_f)}\bigg). \label{pfstochastic}
\end{align}
\end{widetext}
where we neglected the small imaginary part of the sinc$^2$-functions. The summations a, b run over $\omega_1$ and $\omega_2$, and we can assume that $\gamma_{e'} \approx \gamma_e$. In comparison with the excitations induced by entangled photons, we observe that the single-exciton peaks in (\ref{peIIentangled}) and (\ref{Tfgentangled}) have a very similar structure to (\ref{peIIstochastic}) and (\ref{pfstochastic}), respectively. However, eqs. (\ref{peII}) and (\ref{pf}) also contain Lorentzian factors that depend on $\omega_1 + \omega_2$, which dominate the signal. Those resonances are replaced by a constant background contribution in the second lines of eqs. (\ref{peIIstochastic}) and (\ref{pfstochastic}).

\bibliographystyle{apsrev}

\end{document}